\documentclass[twocolumn,aps,showpacs,superscriptaddress]{revtex4}
\usepackage{array}
\usepackage{graphicx}
\usepackage{bm}
\begin{document}

\title{Asymmetry-enriched electronic and optical properties of bilayer graphene}

\author{Bor-Luen Huang}
\affiliation{Department of Physics, National Cheng Kung University, Tainan 701, Taiwan}
\affiliation{Physics Division, National Center for Theoretical Sciences, Hsinchu 300, Taiwan}
\author{Chih-Piao Chuu}
\affiliation{Physics Division, National Center for Theoretical Sciences, Hsinchu 300, Taiwan}
\author{Ming-Fa Lin}
\affiliation{Hierarchical Green-Energy Materials Research Center, National Cheng Kung University, Tainan 701, Taiwan}
\affiliation{Quantum Topology Center, National Cheng Kung University, Tainan 701, Taiwan}

\begin{abstract}

The electronic and optical response of Bernal stacked bilayer graphene with geometry modulation
and gate voltage are studied. The broken symmetry in sublattices, one dimensional periodicity
perpendicular to the domain wall and out-of-plane axis introduces substantial changes of wavefunctions,
such as gapless topological protected states,
standing waves with bonding and anti-bonding characteristics,
rich structures in density of states and optical spectra.
The wavefunctions present well-behaved standing waves in pure system and complicated
node structures in geometry-modulated system.
The optical absorption spectra show forbidden optical excitation channels,
prominent asymmetric absorption peaks, and dramatic variations in absorption structures.
These results provide that the geometry-modulated structure with tunable gate voltage
could be used for electronic and optical manipulation in future graphene-based devices.

\end{abstract}

\date{\today }

\maketitle

\section{Introduction}

The layered structure materials have stimulated enormous studies in the condensed matter
and high-energy physics community.
Since the discovery of single-layer graphene in 2004, \cite{Novoselov04} various 2D materials with
stable planar or buckling structures have been realized, e.g.
mono-element group-IV systems, including
silicene, \cite{Houssa15} germanene,\cite{Davila} and tinene. \cite{Cai15}
One of the exciting findings in graphene-related systems is a new playground
for quantum electromagnetic dynamics.
In fundamental studies and applications,
understanding manipulations on 2D materials become urgent. \cite{CastroNeto09,Bulter13}
For few layers graphene, stacking configuration and external field play critical roles
in diversifying essential physical properties. \cite{Castro08,Castro10}
Bilayer graphenes (BLGs) with AB or AA stacking possess different electronic structures, \cite{Huang14}
absorption spectra, \cite{Chiu03} and Coulomb excitations \cite{Ho06} from each other.
In general, there are different kinds of stacking, including AB, AA, sliding, twisting,
and geometry-modulated BLGs.
On the other side, AB stacking BLG creates a gap when applying an external electric field. \cite{Min07}

Recently, extensive interests focus on the topological characteristics of materials, \cite{Hasan10,Qi11,Ando13}
which have gapless topological protected states
around boundaries and geometry defects.
The domain wall (DW) formed in between by two oppositely biased region of BLG is first proposed by Martin et al. \cite{Martin08}
to host one-dimensional topological states, which is distinct from the edge states in nature. \cite{Castro08prl}
It is believed that this kink of topological defect hosts symmetry-protected gapless mode because of
a change in the Chern number. \cite{Vaezi13,Zhang13} However, experimental realization of such electric-field defined walls
is extremely challenging. An alternative approach is to exploit different stacking orders,
which creates diverse new physics in multi-layer system. \cite{Vaezi13}
Such crystalline topological line defects exist naturally
in Bernal stacked BLGs grown by chemical vapour deposition (CVD) \cite{Alden13,Butz13,Li16}
and in exfoliated BLGs
from graphite. \cite{Ju15,Yin16,Yin17}
This kind of DWs in BLGs can be moved by mechanical stress exerted
through an atomic force microscope (AFM) tip. \cite{Jiang18}
It is also possible to manipulate
DWs with designed structures by controlling the movement of the AFM tip.

The electronic structures and optical absorption show various pictures for different systems.
According to the low-lying electronic structures, AA, AA' (BLG with a special sliding vector),
and AB stackings, respectively, possess vertical and non-vertical two Dirac-cone structures,
and two pairs of parabolic bands. \cite{Lin17}
Density of states (DOS) and optical absorption spectra exhibit 2D van Hove singularities (vHSs),
in which the special structures strongly depend on the characters of relevant dimension.
Similar 2D phenomena appear in sliding and twisted systems, \cite{Huang14,Koshino13,Moon13}
while important differences are revealed between them.
BLG with sliding configurations shows highly distorted energy dispersions
with an eye-shaped stateless region accompanied by saddle points.
In twisted BLGs, the moir\'{e} superlattice enlarges the primitive unit cell
and thus creates a lot of 2D energy subbands.
The geometry-modulated BLG is expected to generate many characters
different from 2D bulk properties. Many works have studied band structures of AB-BA BLG
by effective models \cite{Zhang13,Koshino13,Jiang17}
or tight-binding model with sharp DW. \cite{Vaezi13,Jaskolski18}
Finite but short width has been studied in Ref. \cite{Vaezi13} by first-principle calculation.
However, the width of domain should be large enough to avoid interaction between different DW states.
Reference \cite{Lee16} has studied DW states in large domain and DW width.

In this work, we investigate electronic and optical properties of geometry- and gate-modulated
BLGs in tight-binding model.
We present energy subbands, tight-binding functions on distinct sublattices,
and DOS by exact diagonalization.
Optical absorption spectra are calculated within gradient approximation.
We study the effects caused by geometry modulation and gate voltage.
The band structures show diverse 1D phenomena, including 1D energy subbands with various band-edge states,
band splitting through geometry modulation, and metallic behavior in geometry-modulated system with a gate voltage.
The wavefunctions present well-behaved standing waves in system without geometry modulation,
while their node structures become complicated as applying geometry modulation and/or gate voltage.
We discuss the wavefunctions of the topological protected DW states.
For optical absorption spectra, we find
forbidden optical excitation channels under specific linear relations between layer-dependent sublattices,
prominent asymmetric absorption peaks in absence of selection rule,
and DW- and gate-induced dramatic variations in optical absorption structures.
We also discuss the connection to experiments.
Our predicted results could be verified by the experimental measurements.

This paper is organized as follows. Section \ref{theory} covers the tight-binding
model and the Kubo formula for optical absorption within gradient approximation.
Band structures, tight-binding functions, DOS, joint density of states,
and optical absorption spectra are shown for different bilayer systems in Sec. \ref{electronic}.
The connection to experiments is presented before the end of Sec. \ref{electronic}.
Section \ref{conclusion} is the concluding remarks.

\section{The Theoretical Models}\label{theory}

\begin{figure}
\rotatebox{0}{\includegraphics*[width=9.2cm]{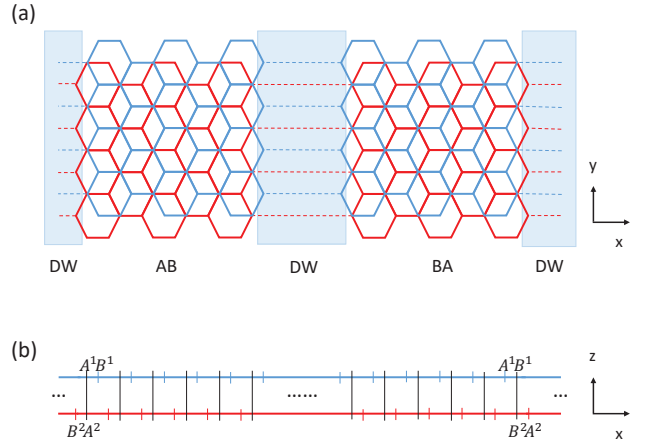}}
\caption{The structure of geometry-modulated (AB-BA) BLG: (a) top view, and (b) side view.
Blue boxes in (a) indicate the regions for DWs.}\label{setup}
\end{figure}

Figure \ref{setup} shows schematic diagram for stacking DWs, also known as stacking solitons,
between AB and BA domains in Bernal-stacked BLG.
Such topological line defects involve both tensile and shear tilt boundaries,
as a consequence of the competition between strain energy and misalignment energy cost
within the DW, spanning in space a few to ten nanometers. \cite{Alden13}
In this paper, we consider tensile soliton, which has zigzag boundary along the DW,
created by uniformly stretching the C-C bonds of one layer along the armchair direction
by an atomic bond distance. We assume the system has translational invariance
along the tangent ($\hat{y}$) direction. For pristine Bernal-stacked BLG,
all carbon atoms in a primitive unite cell are labeled by top- and bottom-layer indices,
and A- and B-sublattice indices, (A$^1$, B$^1$, A$^2$, B$^2$).
The positions of these atoms can further be classified into even and odd for each sublattice.
Because the wavefunctions with even and odd position indices only have a $\pi$ phase difference,
the considered bilayer system will be described by the components of wavefunctions only with odd position indices.
In our notation, A$^1$ and A$^2$ (B$^1$ and B$^2$) are on top of each other for AB (BA) domain, as shown in Fig. \ref{setup}(b).

The low-energy Hamiltonian is described by the single orbital $p_z$ tight-binding model ,
\begin{equation}
H=-\sum_{{\bf r},{\bf r}'}t({\bf r},{\bf r}')[c^{+}({\bf r})c({\bf r}')+h.c.],
\end{equation}
where $c^{+}({\bf r})$ and $c({\bf r})$ are creation and annihilation operators at position ${\bf r}$, respectively.
The hopping amplitude is characterized by the empirical formula \cite{Huang14,Koshino13,Slater54}
\begin{equation}
-t({\bf r},{\bf r}')=\gamma_0 e^{-\frac{d-b_0}{\rho}}(1-(\frac{d_0}{d})^2) + \gamma_1 e^{-\frac{d-d_0}{\rho}} (\frac{d_0}{d})^2,
\end{equation}
where $d=|{\bf r}-{\bf r}'|$ is the distance connecting two lattice points,
$b_0=1.42$ {\AA} the in-plane C-C bond length, $d_0=3.35$ {\AA} the interlayer distance,
and $\rho = 0.184 b_0$ the characteristic decay length.
$\gamma_0 = -2.7$ eV is the intralayer nearest-neighbor hopping integral and
$\gamma_1 = 0.48$ eV is the interlayer interaction.
We consider a system of 100 unite cells for each domain with various width of DWs
($N_{dw}$ strained unit cells with a modified lattice constant $a'=a(3N_{dw}+4)/(3N_{dw}+2)$
for geometry-modulated layer in each DW region, where $a=\sqrt{3}b_0$
is the lattice constant of single layer graphene).
Because of translational invariance along zigzag direction, the quasi-1D Hamiltonian
is a real symmetric matrix, which can possess real eigenfunction for each eigenvalue. \cite{realpsi}
Opposite on-site energy between two layers is included to describe layer-dependent Coulomb potential
to simulate the effect of gate voltage. \cite{Kuzmenko09}

When BLG is subject to an electromagnetic field at zero temperature,
occupied valence states are excited to unoccupied conduction ones by incident photons.
In general, only vertical transitions can occur because of negligible momenta carried by photons.
Based on the Kubo formula for linear response of an electromagnetic field,
the optical absorption function is given by \cite{Lin94,Chang06,Pedersen01,Johnson73,Ehrenreich59}
\begin{equation}
\begin{array}{lr}
A(\omega) \propto &  \sum_{c,v}\int_{\mbox{\tiny 1st BZ}} d^{2}{\bf k}
| \langle \Psi^{c}({\bf k})|\frac{\hat{E}\cdot\vec{P}}{m_e} | \Psi^{v}({\bf k})\rangle |^{2} \\
 & \\ & \times \mbox{Im}\{ \frac{1}{E^{c}({\bf k})-E^{v}({\bf k})-\omega-i\Gamma} \},
\end{array}
\end{equation}
where $\vec{P}$ is the momentum operator, $\hat{E}$ the unit vector of an electric polarization along ${\hat y}$,
$m_e$ the bare electron mass, and $\Gamma$ the broadening factor due to various de-excitation mechanisms.
For a clean sample, physical properties can be observed under a sufficiently low temperature
and $\Gamma$  (=2 meV) will be small enough for observing the fine structures.
The superscripts $c$ and $v$ represent the indices of conduction and valence subbands, respectively.
The velocity matrix element, $\langle \Psi^{c}({\bf k})|\frac{\hat{E}\cdot\vec{P}}{m_e} | \Psi^{v}({\bf k})\rangle$,
for optical properties of carbon-related sp$^2$ bonding
is evaluated from the gradient approximation. \cite{Johnson73,Pedersen01}
The velocity operator is calculated by $\partial H({\bf k})/\partial k_y$.
The $\gamma_0$-dependent velocity matrix elements dominate the optical excitations.
Whether vertical excitation channels could survive is mainly related to
the eigenfunction of the A-sublattice of the initial state to
that of the B-sublattice of the final state within each layer, or vice versa.
$E^{c}({\bf k})-E^{v}({\bf k})$ is the excitation energy.
The joint density of states (JDOS), which is defined by setting the velocity matrix elements
in Eq. (3) equal to one, determine the available number of vertical optical excitation channels.
While JDOS directly links with the band-edge states, it will exhibit the prominent structures.

\section{Electronic and Optical Properties}\label{electronic}

\subsection{Without geometry modulation}

\begin{figure}
\rotatebox{0}{\includegraphics*[width=8.6cm]{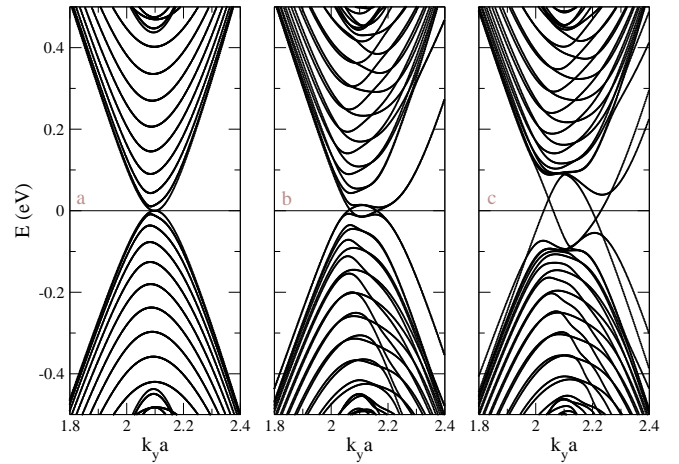}}
\caption{Low-lying energy bands near $\bar{K}$ for (a) AB stacking BLG,
(b) the stacking-modulated system with 20 strained unit cells for each DW,
(c) the system with the same arrangement as (b) in finite gate voltage (0.1 eV).
}\label{energy}
\end{figure}

Low-lying band structures of pure and geometry-modulated
BLGs exhibit unusual features. For pristine AB stacking graphene, the first conduction
and valence bands are 2D parabolic bands and slightly overlap with each other around
$K^{\pm}=(2\pi/\sqrt{3}a,\pm2\pi/3a)$.
We first study the energy bands of AB stacking graphene without geometry modulation
around $\bar{K}$ $(k_y a=2\pi/3)$ in Fig. \ref{energy}(a).
1D parabolic dispersions is due to zone-folding effects.
Because of translational invariance along $x$ direction, each band,
except the first conduction and valence bands, in Fig. \ref{energy}(a) involves two degenerate states,
which can be referred to $\pm k_x$ of the pristine AB stacking graphene. (The first conduction
and valence bands are corresponding to high-symmetry line, therefore they are not degenerate.)
Therefore, the band indices are closed related to discrete $k_x$, provided applying Fourier
transformation along the $x$ direction.
The energy dispersions along $k_x$ is almost negligible due to the sufficiently large unit cell.
The particle-hole symmetry is broken because of non-vanished hopping integrals
between atoms with distance longer than the spacing of the nearest neighbors.
According to state energies measured from $E_F$, the first valence and conduction bands
$(v_1, c_1)$ touch with each other around $\bar{K}$ and have a very weak overlap,
which is consistent with the case in the pristine system.
Their electronic states are non-degenerate except for the spin degrees of freedom.
The other pairs of valence and conduction bands are doubly degenerate and denoted
as $(v_2,c_2)$ and $(v'_2,c'_2)$, $(v_3,c_3)$ and $(v'_3,c'_3)$, and so on.
All band-edge states in distinct energy bands are almost situated at $\bar{K}$ point.
Figure \ref{energy} also shows the second groups of 1D parabolic dispersions, which corresponding to the second
valence and conduction bands of pristine BLG split away from zero energy
by an energy of the order of interlayer coupling $\gamma_1$. \cite{McCann13}

\begin{figure}
\rotatebox{0}{\includegraphics*[width=8.6cm]{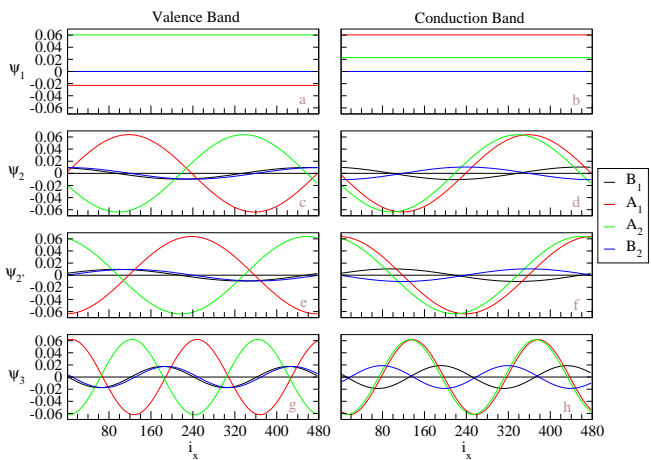}}
\caption{The tight-binding functions at $\bar{K}$ ($k_y a=2\pi/3$) for distinct energy subbands
in BLG without geometry modulation.
For the first valence (a) and conduction (b) bands, the tight-binding functions are non-vanished constant values
for A$^1$- and A$^2$-sublattices. (c-f) show the tight-binding functions for the second valence and conduction
bands has two nodes. (g,h) show
four nodes for each tight-binding function with next larger subband index.
}\label{wf-bulk}
\end{figure}

The wavefunctions at $\bar{K}$, shown in Fig. \ref{wf-bulk}, illustrate main features of
the spatial distributions for various energy subbands.
The eigenfunctions, also known as tight-binding functions, on four sublattices
(A$^1$, B$^1$, A$^2$, B$^2$) are responsible for the four components of the wavefunction.
$v_1$ and $c_1$ states (Figs. \ref{wf-bulk}(a) and \ref{wf-bulk}(b)), which have constant values for each component,
are anti-symmetric and symmetric superpositions of A$^1$- and A$^2$-dependent tight-binding functions, respectively.
Though $v_1$ and $c_1$ have different dominant component in Fig. \ref{wf-bulk}(a),
these two states are degenerate at $\bar{K}$ and can have other ratio between
the weights of A$^1$- and A$^2$-sublattice through linear combination.
Therefore, the wavefunctions at $\bar{K}$ are localized only at A$^1$- and A$^2$-
sublattices, which are on top of each other. Notice that, because of armchair-sharp
in the x direction, the position of atoms can further be classified by even or odd index.
Within the same sublattice, the eigenfunctions with even or odd position index only has a $\pi$ phase difference.
For clear presentation, we only show the components with odd position indices.
With increase of subband indices, the wavefunctions become well-behaved standing waves
instead of uniform spatial distributions.
For $(v_2,c_2)$ or $(v'_2,c'_2)$, the states have dominant components
in the A$^1$- and A$^2$-sublattices and minor weights in B$^1$- and B$^2$-sublattices,
which means that finite momentum away from $\bar{K}$ smear the wavefunction through
finite weights in the sublattices with dangling atoms.
The tight-binding functions show standing-wave behaviors with two nodes for each component
in the spatial distributions, which is due to the linear combination between
$\pm k_x$ as mentioned before and the constrain to have real wavefunctions for a real symmetric matrix. \cite{realpsi}
The $\pi/2$ phase for each sublattice between $v_2$ and $v'_2$ or between $c_2$ and $c'_2$ is consistent
with uniform distribution for the bulk system.
Our results also show that, the eigenfunctions between A$^1$- and A$^2$-sublattices are almost in-phase
for conduction bands and out-of-phase for valence bands,
and those between B$^1$- and B$^2$-sublattices are almost out-of-phase for conduction bands and in-phase for valence bands.
The slight phase shift between A$^1$ and A$^2$ or between
B$^1$- and B$^2$ is intrinsic, depending on the material parameters.
The phase shift becomes less obvious for modes with larger band indices.
For the next band index, the tight-binding functions exhibit four-nodes standing waves
(as shown in Fig. \ref{wf-bulk}(g) and \ref{wf-bulk}(h)). The number of nodes, which exhibits the quantization
behavior in the x-direction, are fixed for all sublattices of each wavefunction
and continuously grow in the increment of subband indices. The detail of wavefunctions would
help to realize the fine structure of optical absorption spectrum.

\subsection{Geometry modulation}

Dramatic changes in electronic structure come to exist for the geometry-modulated BLGs (Fig. \ref{setup}).
Because of more complicated interlayer hopping integrals, asymmetry of the energy spectrum
about Fermi energy is greatly enhanced by the geometric modulation, as clearly displayed in Fig. \ref{energy}(b).
The overlap of valence and conduction bands is getting larger, and so do the free electron and hole densities.
The destruction of the inversion symmetry leads to the splitting of doubly degenerate states.
More pairs of neighboring energy subbands are created.
The splitting electronic states evolve more excitation channels and additional structures in optical absorption spectrum.
Most of the energy bands have parabolic dispersions, while $(v_1,c_1)$ and $(v_2,c_2)$ around half filling
exhibit oscillating and crossing behaviors.
In general, the band-edge states in various energy subbands seriously deviate from $\bar{K}$.
They are responsible for the vHSs in DOS and
thus the number of vertical optical transition channels.

\begin{figure}
\rotatebox{0}{\includegraphics*[width=8.6cm]{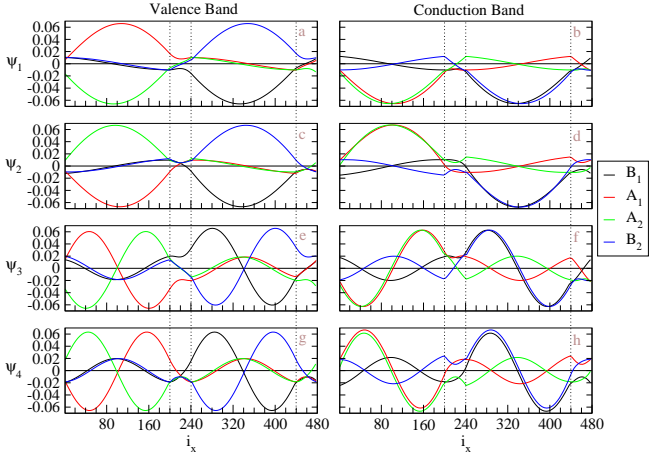}}
\caption{The tight-binding functions at $\bar{K}$ for distinct energy subbands
in the geometry-modulated BLGs. $N_{dw}=20$. The subband index increases from top to bottom.
Each tight-binding function in (a-d) shows two nodes, while that in (e-h) has four nodes.
}\label{wf-dw}
\end{figure}

Spatial distributions of the wavefunctions belong to unusual standing waves.
Symmetric and antisymmetric standing waves for each tight binding function thoroughly
disappear under the modulation of stacking configuration, as clearly indicated in Fig. \ref{wf-dw}.
The tight-binding functions on four sublattices roughly have a linear superposition relationship
within the AB and BA domains.
When the weight is large for one domain, it becomes small for the other.
While the modulation of interlayer hopping amplitudes and relaxed intralayer hopping amplitudes
act as scattering centers for the extended solutions of infinite system,
the weight within DWs is small comparing with that in domains.
Each component of a wavefunction still has fixed number of nodes.
It grows with the 1D subband indices, e.g., two and four nodes in
$(v_1/v'_1,c_1/c'_1)$ and $(v_2/v'_2,c_2/c'_2)$ subbands, respectively.
Wavefunction with zero node disappears because of no translational invariance in the $x$ direction.
The position of nodes can be located within domains or DWs.
The left (right) domain, which corresponding to AB (BA) stacking BLG,
contains the major weight from the eigenfunctions of A$^1$- and A$^2$- (B$^1$- and B$^2$-) sublattices,
consistent with the case without geometry modulation.
For conduction (valence) bands, the eigenfunctions with major weight are in-phase (out-of-phase)
and those with minor weight are out-of-phase (in-phase).
On the other hand, for conduction (valence) bands,
the eigenfunctions between A$^1$ and B$^1$ are symmetric (anti-symmetric) and
anti-symmetric (symmetric) for $c_1$ and $c_2$ ($v_1$ and $v_2$), respectively.
Similar behaviors of the eigenfunctions can be obtained for larger energy subband indices.
Due to the major weight confined within one domain, the phase shift becomes less obvious,
which is similar to the cases with nodes more than 2 in the uniform system.

\subsection{Gate voltage}

Electronic energy spectra are greatly affected by an external electric field.
For pristine AB stacking BLG, the gate voltage ($V_z$) creates a band gap. \cite{Castro10}
Roughly speaking, for AB-BA BLG, each domain still prefers to have a gap when applying a gate voltage.
However, the characteristics of the wavefunctions in the AB domain
is different from that in the BA domain, as discussed in the previous section.
In order to bend in with these two subsystems, the geometry-modulation BLG
with gate voltage will have metallic states within the gap.
The reason is likely to be the two subsystems carry different valley Chern number. \cite{Vaezi13}
Figure \ref{energy}(c) shows rich and unique energy dispersions around the Fermi level.
Most of energy levels repelled away from the Fermi energy by $V_z$,
where the energy dispersions become weak within a certain range of $k_y$.
Moreover, two Dirac dispersions around $\bar{K}$ lead to a finite contribution in DOS and
provide metallic behavior near the Fermi energy,
in great contrast with semiconducting behavior in the gated AB stacking BLGs.

\begin{figure}
\rotatebox{0}{\includegraphics*[width=8.6cm]{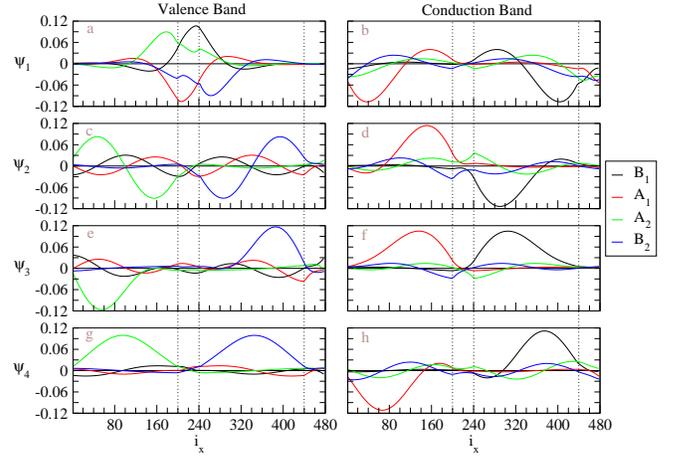}}
\caption{The tight-binding functions at $\bar{K}$ for distinct energy subbands
in the geometry-modulated BLG with a gate voltage, $V_z=0.1$ eV. $N_{dw}=20$.
The subband index increases from top to bottom. The sequence of number of nodes
becomes complicated in the case.
}\label{wf-dwvp1}
\end{figure}

The valence and conduction wavefunctions become highly complicated, as displayed in Fig. \ref{wf-dwvp1},
when the system sustains geometry modulation and gate voltage.
In AB stacking BLG with a positive gate voltage in the $z$ direction, which breaks inversion symmetry,
the tight-binding functions of A$^2$-sublattice are mainly responsible for the valence band states
and those of A$^1$-sublattice for conduction band states.
This is different from the un-gated case, where the top layer and bottom layer components are
distributed evenly between layers, as shown in Fig. \ref{wf-dw}.
In other words, electron states are driven to the top layer, and hole states stay at bottom layer,
and the BLG is therefore polarized under the electric field.
This property of applying gate voltage is still maintained in the geometry-modulation BLGs.
The left (right) column of the plots in Fig. \ref{wf-dwvp1}, which corresponding to conduction (valence) bands,
shows major contribution from the bottom (top) layer except the first row of the plots.
We shall discuss these two states with more details later.
As shown in Fig. \ref{wf-dwvp1}(c-h), the abnormal standing waves show irregular features in oscillatory forms,
amplitudes, numbers of nodes, and relationship among four sublattices.
In general, there are no analytic sine or cosine waves suitable for the spatial distributions of the wavefunctions.
Because the gated AB BLG has the band crossovers in both conduction and valence bands,
the number of nodes does not grow with state energies monotonously.
Furthermore, it might be identical or different for the top and bottom layers.
For example, the $v_1$ ($v_2$) subband at $K$ exhibits the 4- and 4-zero-point (4- and 6-zero-point)
subenvelope functions on the top and bottom layers, respectively (Figs. \ref{wf-dwvp1}(a) and \ref{wf-dwvp1}(c)).
If examining the wavefunctions from higher energy to lower energy (from Fig. \ref{wf-dwvp1} (h) to (b)
of the conduction bands through (a) up to (g) of the valence bands) with fixed $k_y$,
we find that the major weights of the wavefunctions in corresponding domains seem to
inject into the opposite domains through both DWs and have some rebounds.
Similar behavior can be found in the opposite way.
As a result, simple linear combination of the (A$^1$,A$^2$)- and (B$^1$,B$^2$)-related
tight-binding functions disappear. These features of wavefunctions near Fermi energy
are expected to induce very complex optical excitation channels for small energy frequency.

\begin{figure}
\rotatebox{0}{\includegraphics*[width=8.6cm]{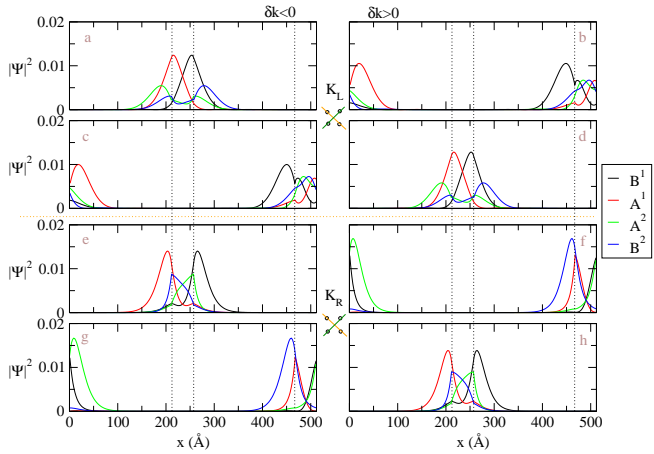}}
\caption{The probability distribution of DW states near $\bar{K}_L$ and $\bar{K}_R$
in the geometry-modulated BLG with a gate voltage (0.1 eV). $N_{dw}=20$.
The real position (along $\hat{x}$) for each component instead of position index is shown in plots.
The orange (green) lines of the schematic diagrams below $K_L$ and $K_R$ represent the linear dispersions
around new Dirac points with negative (positive) group velocities for
the DW states in the middle (another) DW.
}\label{wf-dwvp1-2}
\end{figure}

To compromise different characteristics in different DWs, the major components of the wavefunctions
will leak into the other sublattices with energy outside the corresponding quasi-1D bulk bands.
It is only possible to create this non-uniform distribution near DWs.
A pair of special wavefunctions is shown in Fig. \ref{wf-dwvp1}(a)-(b),
which has significant enhancement for the weight within the DWs.
These DW states are robust once the system has a gate voltage.
The localization in Fig. \ref{wf-dwvp1}(b) is weak, where the energy level at $\bar{K}$
is close to the band edges created by the gate voltage.
When the $k_y$ is set to be near the crossing points ($\bar{K}_L$ and $\bar{K}_R$),
the DW states become apparent, as shown in Fig. \ref{wf-dwvp1-2}.
In Fig. \ref{wf-dwvp1-2}, instead of tight-binding functions,
we present the probability distribution for each component in real space,
which can be directly measured in experiments (discussed later).
Note that, even we name these localized states DW states, the weight outside
DW is still large and the corresponding decay length is around several nanometers for $V_z=0.1$ eV.
The decay length becomes smaller as $V_z$ is larger.
Therefore, large size of domains, as considered in this paper, is crucial
to have well-define DW states when the applied gate voltage is limited.
Two DW states, localized at each DW, belong to two
different linear dispersive bands but with group velocities of the same sign.
The DW states localized at the first DW have negative group velocities in $\hat{y}$
and dominant weights for top (bottom) layer when $k$ is around $K_L$ ($K_R$),
as shown in Fig. \ref{wf-dwvp1-2} (a,d) (Fig. \ref{wf-dwvp1-2} (e,h)).
On the other hand, the DW states localized at the second DW have positive group velocities in $\hat{y}$
and dominant weights for bottom (top) layer when $k$ is around $K_L$ ($K_R$),
as shown in Fig. \ref{wf-dwvp1-2} (b,c) (Fig. \ref{wf-dwvp1-2} (f,g)).
The dominant component of these DW states can be different from that of the other states
which may be referred to bulk states in this gated geometry-modulated system.
For example, probability distribution of A$^1$ component of the DW states in
Fig. \ref{wf-dwvp1-2} (f) and (g) appears in the right (BA) domain with countable weight in the DW,
while that of bulk states is used to be located at the left (AB) domain.
These unusual properties of DW states, which is presented in real space wavefunctions,
might be used to distinguish topological protected states from the others.

\subsection{Density of state}

\begin{figure}
\rotatebox{0}{\includegraphics*[width=8.6cm]{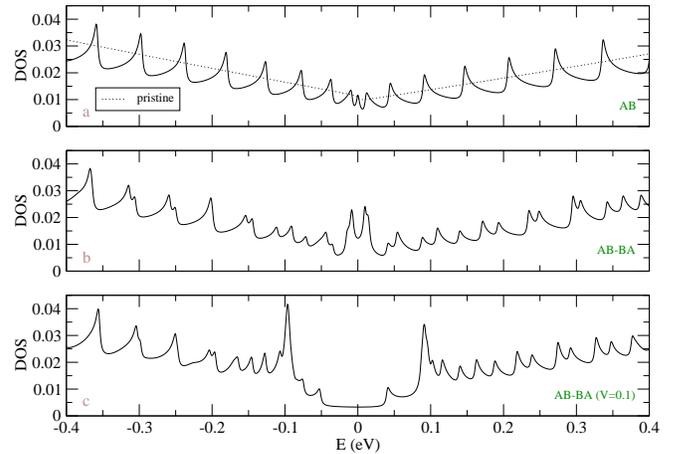}}
\caption{Density of states for AB stacking BLG (a) without geometry modulation,
(b) with geometry modulation ($N_{dw}=20$.), and (c) the later with finite gate voltage ($V_z=0.1$ eV).
$\Gamma=2$meV is small enough for observing the fine structures.}\label{dos}
\end{figure}

Main features of electronic structures are directly reflected in the DOS, as shown in Fig. \ref{dos}.
The pristine system has a finite DOS at $E_F$ (the dot line in Fig. \ref{dos}(a)), indicating the gapless property.
Weak overlap between conduction and valence bands leads to a very small shoulder structure at $E_F$.
In addition to a pair of shoulder structures around $\gamma_1$, which comes from the valence and conduction bands
of the second group, the absence of special structures near $E_F$ is consistent with
the first-principles calculations. \cite{McCann13} In this work, we focus on the fine structure
within small energy region and do not show the shoulder structures around $\gamma_1$ in Fig. \ref{dos}.
For a finite system without geometry modulation, Fig. \ref{dos}(a) presents a lot of asymmetric peaks of DOS
divergent in square-root form, due to 1D parabolic energy subbands as shown in Fig. \ref{energy}(a).
On the other side, the geometry modulation can create additional peaks (Fig. \ref{dos}(b-c)), which corresponds
to the splitting of doubly degenerate states (Fig. \ref{energy}(b-c)).
A finite DOS at $E_F$ in the geometry-modulated system shows metallic behavior,
combined with a pair of rather strong peak structures around $\pm 0.01$ eV.
The latter are related to the weakly dispersive energy bands near $E_F$.
With a gate voltage, the DOS near $E_F$ becomes a plateau structure, which is
created by the linear energy dispersions across $E_F$, as shown in Fig. \ref{energy}(c).
Two very prominent peaks appear around $\pm V_z$ mainly due to the weak dispersions generated by the gate voltage.
There are also some sub-structures for $|E|<V_z$, which corresponds to
band edge states with major weights around DWs. These accidental sub-structures
come from large size of DWs.

\subsection{Optical absorption}

\begin{figure}
\rotatebox{0}{\includegraphics*[width=8.6cm]{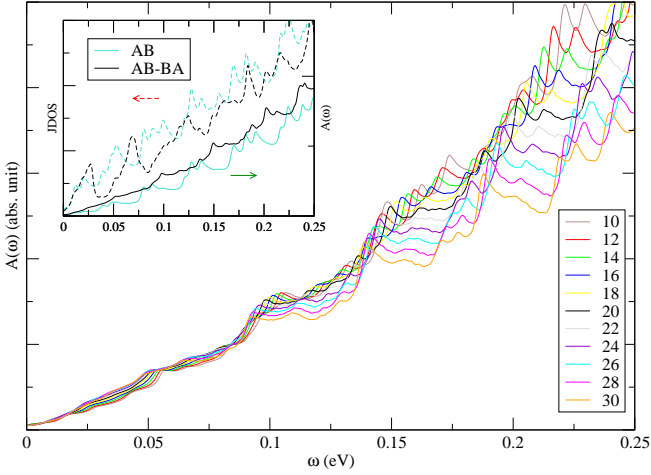}}
\caption{The optical absorption under the various DW widths (marked by $N_{dw}$).
Also shown in the inset are JDOS and ${A(\omega)}$ in BLG
without and with the geometry modulation ($N_{dw}=20$).}\label{oj-dw}
\end{figure}

\begin{figure}
\rotatebox{0}{\includegraphics*[width=8.6cm]{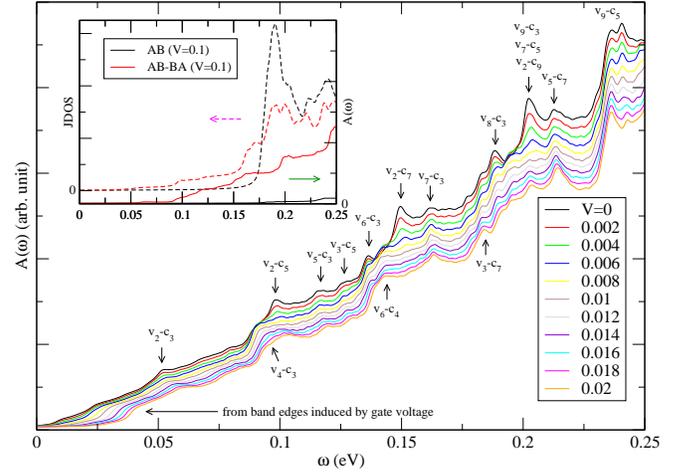}}
\caption{The optical absorption for a specific geometry-manipulated BLG
in presence of distinct gate voltages. $N_{dw}=20$. The arrows indicate the special channels
for the detailed absorption structures. The inset shows JDOS and ${A(\omega)}$
in BLG without and with the geometry modulation for $V_z=0.1$ eV.  }\label{oj-dv}
\end{figure}

The geometry- and gate-modulated BLG exhibits rich and unique optical properties.
The JDOS gives the weight of vertical optical excitations between different bands.
For pristine AB stacking BLG, JDOS shows monotonically increasing
within the considered frequency region, \cite{Lu06} which has the same reason to have monotonically
increasing in DOS for small energy region away from $E_F$.
The inset of Fig. \ref{oj-dw} shows peak structures, which indicate many channels between
different 1D subbands, in the JDOS of BLG without and with geometry modulation (the dashed curves).
Peaks in the JDOS are strong when the valence and conduction band-edge states have the same wave vector,
e.g. $(v_1,c_1,v_2,c_2)$ around $\bar{K}$ and $(v_2,c_3)$ at $k_y a\approx2.06$ for BLG with DWs.
There are some weak but observable structures due to non-vertical relation between different subbands.
The peak structures in the JDOS are greatly reduced when system is under a gate voltage,
as shown in the inset of Fig. \ref{oj-dv}, because most states with $|E|<V_z$ are repelled away from Fermi energy
(Fig. \ref{energy}(c)). For the system without geometry modulation, the threshold vertical
channels is around $V_z=0.1$ eV. However, some weak structures appear in the geometry-modulated BLG
for $\omega<V_z$
because of linear dispersions and band edge states with major weights around DWs (Fig. \ref{energy}(c)).
This is one of major differences in optical absorption when geometry modulation is included.

Structures in the optical absorption spectrum is determined by
the available optical transition channels and the velocity matrix elements between channels.
From Fig. \ref{oj-dw}, we can find both optical gap and energy gap are identical and equal to zero.
The optical absorption spectrum also shows 1D absorption peaks.
However, many peaks in JDOS do not appear in the optical absorption spectrum,
as shown in the insets of Fig. \ref{oj-dw} and \ref{oj-dv}.
The relations between wavefunctions of valence and conduction bands
strongly affect the appearance of absorption peaks.
The $v_n$ to $c_n$ vertical excitations, which are due to the same pair of valence and conduction bands, are forbidden.
The main mechanism is the linear symmetry or anti-symmetric superposition of
the $(A^1,A^2)$ and $(B^1,B^2)$ sublattices, as discussed in Fig. \ref{wf-bulk}.
When the geometry modulation is introduced, the peaks are weaker as compared to uniform system
and the curve for the optical absorption spectrum becomes smoother.
The intensity, frequency, and structures of optical special absorption are
very sensitive to the changes in the width of DW (Fig. \ref{oj-dw})
because of the shifting of band-edge states of the valence and conduction subbands.
We find red-shift phenomena when increasing width of DW and suppressed strength for large $\omega$.
The first effect is mainly due to more atoms involved in the calculation and more energy subbands as a result.
The second effect is related to larger band splitting for larger width of DW.
Also, the first absorption structure might be replaced by
another excitation channel during the variation of DW width.

The changes of the structures of optical absorption spectra become clear when
applying various gate voltages, as shown in Fig. \ref{oj-dv}.
The reduced intensity and the enhanced number of absorption structures occur in the increment of $V_z$.
This directly reflects gradual changes in energy dispersions,
band-edge states and wavefunctions with the gate voltages.
The first absorption peak, corresponding to $(v_2,c_3)$ at $k_y a=2.06$, for $V_z=0$ appears at 0.049 eV.
First few absorption peaks are indicated by arrows above the curves
with corresponding absorption channels in Fig. \ref{oj-dv}.
By comparing the band structures and corresponding optical absorption spectra carefully,
we find that those peaks involving $v_2$ (the corresponding band edges are sensitive to $V_z$)
are shifted to higher frequency and become weaker when the gate voltage becomes larger.
We find strong peaks which are contributed from multi--optical excitation channels.
Some weak absorption structures, which corresponds to new optical excitation channels, emerge
in between as indicated by arrows below the curves in Fig. \ref{oj-dv}.
New shoulder structures for frequency around $2V_z$ become clearer for lager $V_z$
because of the band edges induced by the gate voltage.
We note that the optical absorptions get greatly enhanced when the system has geometry modulation,
as shown in the inset of Fig. \ref{oj-dv}. Most of the contribution is from
the linear dispersions and band edge states with major weights around DWs (Fig. \ref{energy}(c)).

\subsection{Connection to experiments}

The periodic boundary condition in asymmetry-enriched BLGs is responsible
for 1D electronic and optical properties.
The energy dispersions might belong to linear, parabolic and oscillatory forms, in which the first ones are
the well-known vertical/non-vertical Dirac cones in the AA/AA$^\prime$ stacking. \cite{Huang14,Thuy15,Lin17}
The band-edge states, which are the critical points in energy dispersion,
correspond to 
the extreme points, saddle points, and effective 1D constant-energy loops.
DOS presents various structures, including V-shape structures from linear energy dispersions,
shoulder structures from 2D band edges, logarithmically symmetric peaks from the vHSs,
and square-root asymmetric peaks from 1D band edges. These four kinds of special structures can
further be revealed in the optical absorption spectra. \cite{Lin18}
An electric field would create optical and energy gaps in most of BLG systems. \cite{Castro08,Castro10,Min07}

The experimental measurements can verify the predicted band structures,
DOS, wavefunctions, and optical absorption spectra.
The high-resolution angle-resolved photoemission spectroscopy (ARPES) can be used to
directly examine the energy dispersion.
The measured results have confirmed feature-rich band structures of the carbon-related sp$^2$-bonding systems.
Graphene nanoribbons are identified to possess 1D parabolic energy subbands centered at the high-symmetry point,
accompanied with an energy gap and non-uniform energy spacings. \cite{Ruffieux12}
Recently, a lot of ARPES measurements are conducted on few-layer graphenes,
covering the linear Dirac cone in monolayer system, \cite{Ohta07,Siegel13,Bostwick07}
two pairs of parabolic bands in AB stacking BLG, \cite{Ohta06,Ohta07}
the coexistent linear and parabolic dispersions in symmetry-destroyed bilayer systems, \cite{Kim13}
one linear and two parabolic bands in tri-layer ABA stacking, \cite{Ohta07,Coletti13} the linear,
partially flat and sombrero-shaped bands in tri-layer ABC stacking. \cite{Coletti13}
The Bernal-stacked graphite possesses 3D band structure, with the bilayer- and monolayer-like
energy dispersions, respectively, at K and H points of the first Brillouin zone.
The ARPES examinations on the geometry-modulated BLGs could provide unusual band structures,
such as, the split of electronic states, diverse energy dispersions near $E_F$,
band-edge states, and metallic properties. These directly reflect the strong competition
between stacking symmetries, interlayer hopping integrals and Coulomb potential.

The scanning tunneling spectroscopy (STS) measurements, in which the differential conductance (${dI/dV}$)
is proportional to DOS, are very powerful in exploring the vHSs due to the band-edge states
and the metallic/semiconducting/semi-metallic behaviors. They have successfully identified diverse
electronic properties in graphene nanoribbons, \cite{Huang12,Sode15,Chen13} carbon nanotubes, \cite{Wilder98,Odom98}
few-layer graphenes, \cite{Que15,Luican11,Li10,Cherkez15,Lauffer08,Yankowitz13,Pierucci15} and
graphite. \cite{Klusek99,Li09}
Concerning graphene nanoribbons, the width- and edge-dominated energy gaps and the asymmetric peaks due to 1D parabolic bands
are confirmed from the precisely defined crystal structures. \cite{Magda14,Wilder98,Odom98,Huang12,Sode15,Chen13}
The similar prominent peaks obviously appear in carbon nanotubes, where they present chirality-
and radius-dependent band gaps and energy spacings between two neighboring subbands. \cite{Wilder98,Odom98}
A plenty of STS measurements
on few-layer graphenes clearly reveal diverse low-lying DOS, including a V-shape dependence
initiated from the Dirac point in monolayer system, \cite{Li09} the peak structures closely related to
saddle points in asymetric BLGs, \cite{Luican11,Li10,Cherkez15} an electric-field-induced gap
in bilayer AB stacking and tri-layer ABC stacking, \cite{Lauffer08,Yankowitz13} a pronounced peak at $E_F$
due to partially flat bands in tri-layer and penta-layer ABC stackings, \cite{Wilder98,Odom98} and
a dip structure at $E_F$ accompanied with a pair of asymmetric peaks arising from
constant-energy loops in tri-layer AAB stacking. \cite{Wilder98}
The measured DOS of the AB-stacked graphite is finite near $E_F$ characteristic of semi-metallic property \cite{Li09}
and exhibits the splitting $\pi$ and $\pi^\ast$  strong peaks at deeper/higher energy. \cite{Klusek99}
The focuses of the STS examinations on the geometry-modulated \cite{Li16} and gated BLGs should be
the square-root asymmetric peaks, the single- or double-peak structures,
finite DOS at $E_F$, and strong valence and conduction peaks caused by gate voltage.
The STS can also be used to measure the two-dimensional structure of individual wavefunctions
in metallic single-walled carbon nanotubes. \cite{Venema99,Lemay01}
Therefore, the wavefunctions found in the paper can be verified by STS measurements.

Up to date, four kinds of optical spectroscopies, including absorption, transmission, reflection,
and Raman scattering spectroscopies, are frequently utilized to accurately explore vertical optical excitations. \cite{Lin18}
Concerning the AB-stacked BLG, the experiments have confirmed the 0.3-0.4 eV
shoulder structure under zero field, the $V_z$-created semimetal-semiconductor transition and
two low-frequency asymmetric peaks, the two strong $\pi$-electronic absorption peaks at the middle frequency,
specific magneto-optical selection rule for the first group of Landau levels (LLs),
and linear magnetic-field-strength dependence of the inter-LL excitation energies.
Similar verifications performed on trilayer ABA stacking cover one shoulder around 0.5 eV,
the gapless behavior unaffected by gate voltage, the $V_z$-induced low-frequency multi-peak structures,
several $\pi$-electronic absorption peaks, and monolayer- and bilayer-like inter-LL absorption frequencies.
Moreover, the identified spectral features in trilayer ABC stacking are two low-frequency characteristic peaks
and gap opening under an electric field. The above-mentioned optical spectroscopies are available
in examining the vanishing optical gaps for any metallic systems, prominent asymmetric absorption peaks,
absence of selection rule, forbidden optical excitations associated with linear relations
in the (A$^1$,A$^2$) and (B$^1$,B$^2$) sublattices, and the variations in
absorption structures due to the modulation of DW width and gate voltage.

\section{Concluding Remarks}\label{conclusion}

We have studied the electronic and optical properties of AB stacking BLG with
geometry modulation and gate voltage in tight-binding model.
We present energy subbands, tight-binding functions on distinct sublattices,
and DOS by exact diagonalization.
Effects of geometry modulation in the presence or absence of a gate voltage are discussed.
The metallic systems exhibits a plenty of 1D energy subbands,
accompanied by well-behaved or irregular standing waves.
Specifically, the layer-dependent Coulomb potential destroys a simple relation
between the (A$^1$,A$^2$) and (B$^1$,B$^2$) subenvelope functions and gives
complicated node structures.
With a gate voltage, the system is still metallic due to the existence of the DW states.
The wavefunctions of the topological protected DW states present unusual space distributions.
DOS shows various vHSs, including single- and double-peak structures with the square-root divergent forms,
a pair of prominent peaks caused by gate voltage, and a plateau structure across $E_F$.

Optical absorption spectra are calculated within gradient approximation.
We find forbidden optical excitation channels under specific linear relations
between layer-dependent sublattices, prominent asymmetric absorption peaks in absence of selection rule,
and DW- and gate-induced dramatic variations in optical absorption structures.
Concerning the geometry-modulated systems, the observable absorption peaks could
survive only under the destruction of symmetric or anti-symmetric linear superposition
of the (A$^1$,A$^2$) and (B$^1$,B$^2$) sublattices.
Simple dependence of absorption structures on the modulation width of DW is absent.
However, the reduced intensity and the enhanced number are regularly revealed in the increment of $V_z$.
The frequency, number, and form intensity of optical absorption peaks strongly depend on
the modulation period and electric-field strength.
Our predicted results could be verified by the experimental measurements.

This geometry-modulated system is suitable for studying various physical phenomena.
For example, magneto-electronic properties, which exist in a uniform
perpendicular magnetic field, might be dominated by the 1D Landau subbands,
being in sharp contrast with the Landau levels in 2D systems.
This problem is under current investigation.
How to generalize the manipulations of geometric structures to emergent layered materials
is the near-future focus, e.g., important studies on essential properties
of bilayer silicene, \cite{Padilha15} germanene, \cite{Maria16} phosphorene, \cite{Dai14} and bismuthene. \cite{Ast03}

BLH acknowledge the Center for Micro$/$Nano Science and Technology
for their generous support.
This work was supported in part by the National Science Council of Taiwan under
grant number NSC 105-2112-M-006-002-MY3.


\begin{thebibliography}{plain}
\bibitem{realpsi} If $H\psi=E\psi$, $H\psi^*=E\psi^*$ with real $E$ for real symmetric $H$.
When $H$ does not have degeneracy, $\psi=\psi^*\in$ real. When $H$ have degeneracy,
two of eigenfunctions can be chosen as $\psi+\psi^*\in$ real and $i(\psi-\psi^*)\in$ real.
Therefore, all eigenfunctions of $H$ can be real.
\end{thebibliography}
\end{document}